\documentclass{ws-procs975x65}

\begin{document}

\title{Second-order gravitational self-force---a quick summary}

\author{Adam Pound$^*$}

\address{School of Mathematics, University of Southampton,\\
Southampton, SO17 1BJ, UK\\
$^*$E-mail: A.Pound@soton.ac.uk\\
www.southampton.ac.uk}

\begin{abstract}
In order to extract physical parameters from the waveform of an extreme-mass-ratio binary, one requires a second-order--accurate description of the motion of the smaller of the two objects in the binary. Using a method of matched asymptotic expansions, I derive the second-order equation of motion of a small, nearly spherical and non-rotating compact object in an arbitrary vacuum spacetime. I find that the motion is geodesic in a certain locally defined effective metric satisfying the vacuum Einstein equation through second order, and I outline a method of numerically calculating this effective metric.
\end{abstract}


\bodymatter
\section{Introduction}

Recent years have seen substantial progress in describing compact binary inspirals using a self-force model, in which the smaller object in the binary is treated as a source of perturbation of the spacetime of the larger object.\cite{Poisson-Pound-Vega:11, Barack:09} However, nearly all this progress has been made at the level of linearized gravity. Scaling arguments show that in order to accurately model an inspiral over a radiation-reaction time (the timescale on which orbital parameters change significantly), we require an equation of motion accurate through second order in the smaller object's mass $m$. This requires finding the second-order perturbation produced by the object. 

Here, I summarize recent work\cite{Pound:12a,Pound:12b} that derived the essential analytical ingredients necessary to numerically evolve a second-order--accurate orbit in a hyperbolic region of an arbitrary vacuum background. These ingredients are (i) a local expansion of the metric around the small object, (ii) an equation of motion in terms of certain pieces of that local expansion, and (iii) a means of numerically calculating those certain pieces. A simulation of a binary inspiral can be performed after specializing these three ingredients to the Kerr background of the larger object.


\section{Metric around the small object}\label{metric}
To find a local expansion of the metric, I solve the EFE in a vacuum region \emph{outside} the object. Let ${\sf g}_{\mu\nu}$ be the exact metric, $g_{\mu\nu}$ be a vacuum background metric, $h_{\mu\nu}\equiv{\sf g}_{\mu\nu}-g_{\mu\nu}$ be the field due to the object, $z^\mu(\tau,m)$ be a curve that will represent the object's mean motion in the spacetime of $g_{\mu\nu}$, and $\mathcal{R}$ be the smallest lengthscale of $g_{\mu\nu}$. I construct an expansion of $h_{\mu\nu}$ for small $m$ and $r$ in a ``buffer region" $m\ll r\ll \mathcal{R}$ around the object, where $r$ is a measure of spatial distance from $z^\mu$.



After adopting the Lorenz gauge condition $\nabla^\mu(h_{\mu\nu}-\frac{1}{2}g_{\mu\nu}g^{\rho\sigma}h_{\rho\sigma})=0$ and expanding $h_{\mu\nu}$ as $h^{(1)}_{\mu\nu}+h^{(2)}_{\mu\nu}+O(m^3)$, where $h^{(n)}_{\mu\nu}=O(m^n)$, I write the first- and second-order Einstein equations as the wave equations
\begin{equation}
E_{\mu\nu}{}^{\rho\sigma}h^{(1)}_{\rho\sigma}= 0, \qquad E_{\mu\nu}{}^{\rho\sigma}h^{(2)}_{\rho\sigma}= 2\delta^2 R_{\mu\nu}[h^{(1)}],\label{EFE}
\end{equation}
where $E_{\mu\nu}{}^{\rho\sigma}$ is the tensor wave operator
\begin{equation*}
E_{\mu\nu}{}^{\rho\sigma}h_{\rho\sigma}\equiv \Box h_{\mu\nu}+2R_\mu{}^\rho{}_\nu{}^\sigma h_{\rho\sigma},
\end{equation*}
and $\delta^2 R_{\mu\nu}$ is the second-order Ricci tensor, quadratic in $h^{(1)}_{\mu\nu}$.

The solutions to Eq.~\eqref{EFE} must be compatible with the presence of a compact object in the region $r\lesssim m$, which restricts them to the form $h^{(1)}_{\mu\nu}\sim 1/r+O(r^0)$ and $h^{(2)}_{\mu\nu}\sim 1/r^2+O(1/r)$, a fact derived from the method of matched asymptotic expansions\cite{Pound:10a,Pound:12b}. After imposing this restriction, the remaining freedom in the general solution is found by decomposing the field into spherical harmonics around the object. The freedom consists of a set of functions $h^{(2\ell\pm)}_{\mu\nu}$ that appear in the homogeneous wave equation's flat-space solutions, the harmonic modes of which have the familiar form $h^{(2\ell-)}_{\mu\nu}(t-r)/r^{\ell+1}$ and $h^{(2\ell+)}_{\mu\nu}(t-r)r^\ell$; the functions $h^{(2\ell-)}_{\mu\nu}$ correspond to the object's multipole moments (or corrections thereto), while $h^{(2\ell+)}_{\mu\nu}$ corresponds to free radiation. Using these results, I split the physical field as $h^{(n)}_{\mu\nu}=h^{{\rm S}(n)}_{\mu\nu}+h^{{\rm R}(n)}_{\mu\nu}$, where the \emph{singular field} $h^{{\rm S}(n)}_{\mu\nu}\sim 1/r^n$ comprises all the terms in $h^{(n)}_{\mu\nu}$ involving the functions $h^{(2\ell-)}_{\mu\nu}$, and the \emph{regular field} $h^{{\rm R}(n)}_{\mu\nu}\sim r^0$ comprises all the terms involving only the functions $h^{(2\ell+)}_{\mu\nu}$. This split is analogous to the Detweiler-Whiting decomposition of $h^{(1)}_{\mu\nu}$:\cite{Detweiler-Whiting:02} $h^{{\rm S}(n)}_{\mu\nu}$ characterizes the object's bound field, while $h^{{\rm R}}_{\mu\nu}=h^{{\rm R}(1)}_{\mu\nu}+h^{{\rm R}(2)}_{\mu\nu}$ is a smooth solution to the vacuum EFE through second order. I specialize to a nearly spherical object by setting to zero all the object's moments but the monopole. 

\section{Second-order equation of motion}\label{motion}
I wish $z^\mu$ to represent the object's center of energy. To make it do so, I insist that the local expansion in the Lorenz gauge is equal, up to a gauge transformation, to a local expansion in which the object is manifestly at rest and centered on $z^\mu$. In that `rest gauge', the metric looks like tidally perturbed Schwarzschild. To ensure that the transformation to this other gauge does not move the object's position relative to $z^\mu$, I also insist that it involves no finite translations at $z^\mu$. These conditions together determine $z^\mu$'s acceleration in the Lorenz gauge to be 
\begin{equation}
\frac{D^2 z^\mu}{d\tau^2} = -\frac{1}{2}\left(g^{\mu\nu}+u^\mu u^\nu\right)\left(g_\nu{}^\rho-h^{\rm R}_\nu{}^\rho\right)
		\left(2h^{\rm R}_{\rho\sigma;\lambda}-h^{\rm R}_{\sigma\lambda;\rho}\right)u^\sigma u^\lambda+O(m^3),\label{force}
\end{equation}
which, through second order, is the geodesic equation in the smooth metric $g_{\mu\nu}+h^{\rm R}_{\mu\nu}$.

Similar methods have also been used by Rosenthal\cite{Rosenthal:06b} and Gralla\cite{Gralla:12} to derive second-order self-force equations. In Rosenthal's work, the end result is in an impractical gauge in which the first-order force vanishes. In Gralla's work, the object's worldline is expanded as $z^\mu(\tau,m)=z^\mu_0(\tau)+m z^\mu_1(\tau)+m^2 z^\mu_2(\tau)+O(m^3)$, and a gauge is constructed in which the object is at rest and centered on the zeroth-order curve $z^\mu_0(\tau)$; when transforming to another gauge, translations at $z^\mu_0$, rather than being required to vanish, correspond to the deviations $z^\mu_n$. The disadvantage to that approach is that it is limited to short timescales, since the object will deviate far from $z_0^\mu$ over a radiation-reaction time.

\section{Puncture scheme}\label{puncture}
The local analysis yields an equation of motion in terms of a regular field in the neighbourhood of the object, but it does not actually determine the value of that regular field. Nor can it determine the distant waveforms produced by the object's motion. To calculate these two fields, we can use a numerical puncture scheme,\cite{Barack-Golbourn:07,Vega-Detweiler:07} which replaces the physical problem in and around the object with an effective problem. Define the \emph{puncture} $h^{\mathcal{P}(n)}_{\mu\nu}$ to be $h^{{\rm S}(n)}_{\mu\nu}$ truncated at order $r$ (or higher). Now surround the object by a timelike tube $\Gamma$ and define the \emph{effective field} $h^{{\rm eff}(n)}_{\mu\nu}$ to be identical to the retarded field $h^{(n)}_{\mu\nu}$ outside $\Gamma$, and to be given by the difference $h^{(n)}_{\mu\nu}-h^{\mathcal{P}(n)}_{\mu\nu}$ inside $\Gamma$. This construction ensures that $h^{{\rm eff}(n)}_{\mu\nu}$ agrees with the regular field $h^{{\rm R}(n)}_{\mu\nu}$ through order $r$ near $z^\mu$, meaning it can be used in the equation of motion~\eqref{force}. Rewriting Eq.~\eqref{EFE} in terms of $h^{{\rm eff}(n)}_{\mu\nu}$ and $h^{\mathcal{P}(n)}_{\mu\nu}$ yields the wave equations
\begin{align*}
E_{\mu\nu}{}^{\rho\sigma}h^{{\rm eff}(1)}_{\rho\sigma} &= \begin{cases} -16\pi \bar T^{(1)}_{\mu\nu}-E_{\mu\nu}{}^{\rho\sigma}h^{\mathcal{P}(1)}_{\rho\sigma}  & \text{inside } \Gamma \\
  0 &\text{outside } \Gamma,\end{cases}\\
E_{\mu\nu}{}^{\rho\sigma}h^{{\rm eff}(2)}_{\rho\sigma} 
			&= \begin{cases}-16\pi \bar T^{(2)}_{\mu\nu}+2\delta^2R_{\mu\nu}[h^{{\rm eff}(1)}+h^{\mathcal{P}(1)}]
			-E_{\mu\nu}{}^{\rho\sigma}h^{\mathcal{P}(2)}_{\rho\sigma} &\text{inside }\Gamma\\
			2\delta^2R_{\mu\nu}[h^{{\rm eff}(1)}] &\text{outside }\Gamma,\end{cases}
\end{align*}
where $T^{(n)}_{\mu\nu}$ is an effective stress-energy for the object, defined by distributional singularities in $E_{\mu\nu}{}^{\rho\sigma}h^{\mathcal{P}(n)}_{\rho\sigma}$, and an overbar denotes trace-reversal with $g_{\mu\nu}$. On the right-hand side, all quantities are analytically known function(al)s of $m$ and $h^{{\rm eff}(1)}_{\mu\nu}$. The singularities in these functions cancel one another, leaving regular sources.

In this scheme, the punctures $h^{\mathcal{P}(n)}_{\rho\sigma}$ diverge on the worldline $z^\mu$ that satisfies Eq.~\eqref{force}. Therefore, the wave equations and the equation of motion must be solved self-consistently, as a coupled system. This contrasts with the second-order formalism of Gralla;\cite{Gralla:12} there, the expansion of the worldline leads to punctures that diverge on the zeroth-order curve $z_0^\mu$, and $z_0^\mu$ may be specified in advance of solving the wave equations.

\bibliographystyle{ws-procs975x65}
\bibliography{main}

\end{document}